\documentclass{osa-article}

\journal{osajournal}

\articletype{Research Article}

\begin{document}

\title{The effect of plasma-core induced self-guiding on phase matching of high-order harmonic generation in gases}

\author{Bal\'azs Major,\authormark{1,2,*} Katalin Kov\'acs,\authormark{3} Valer Tosa,\authormark{3} Piotr Rudawski,\authormark{4} Anne L'Huillier,\authormark{4}  and Katalin Varj\'u\authormark{1,2}}

\address{\authormark{1}ELI-ALPS, ELI-HU Non-Profit Ltd., Dugonics ter 13, Szeged 6720, Hungary\\
\authormark{2}Department of Optics and Quantum Electronics, University of Szeged, D\'om t\'er 9., Szeged 6720, Hungary\\
\authormark{3}National Institute for R\&D of Isotopic and Molecular Technologies, 67-103 Donat str., 400293 Cluj-Napoca, Romania\\
\authormark{4}Department of Physics, Lund University, P. O. Box 118, SE-22100 Lund, Sweden
}

\email{\authormark{*}Balazs.Major@eli-alps.hu} 



\begin{abstract}
In this work we numerically study a self-guiding process in which ionization plays a dominant role and analyze its effect on high-order harmonic generation (HHG) in gases. Although this type of self-guiding --- termed as plasma-core induced self-guiding in previous works --- limits the achievable cut-off by regulating the intensity of the laser beam, it provides favorable conditions for phase matching, which is indispensable for high-flux gas high-harmonic sources. To underline the role of self-guiding in efficient HHG, we investigate the time-dependent phase matching conditions in the guided beam and show how the spatio-temporally constant fundamental intensity contributes to the constructive build-up of the harmonic field in a broad photon-energy range up to the provided cut-off.
\end{abstract}

\section{Introduction}

Several emerging research fields in ultrafast physics \cite{Krausz2014NPhot} and chemistry \cite{Gallmann2012ARPC,Lepine2014NPhot, Ramasesha2016ARPC} 
rely on sources based on high-order harmonic generation (HHG).
HHG is most often carried out by focusing an energetic ultrashort laser pulse into an ensemble of noble gas atoms \cite{McPherson1987JOSAB, Ferray1988JPB, Chatziathanasiou2017Photonics}. This interaction provides pulses with duration in the attosecond domain \cite{Midorikawa2011JJAP, Sansone2011NPhot, Chini2014NPhot}, which is an important advantage of HHG over other sources of coherent extreme-ultraviolet
(XUV) and soft x-ray radiation \cite{CanovaPoletto}.
HHG in gases, however, has a low efficiency, so it is very important to consider the 
macroscopic aspects of the generation process \cite{Gaarde2008JPB}, i.e., to apply conditions
where the radiation can be generated phase matched.
Phase matching is strongly influenced by the way the
intensity and phase of the generating laser and high-harmonic fields change during propagation through the interaction region \cite{Heyl2016JPB}. 

It has been well-known for decades that when the power of a pulsed laser beam exceeds a critical value, self-trapping and long-distance guiding of the laser beam can happen \cite{Chiao1964PRL,Moll2003PRL}. This mechanism is attributed to the balance between diffraction, Kerr self-focusing and plasma defocusing \cite{Esarey1997IEEEJQE}. Self-guiding is more easily achievable with atmospheric or high pressure gases \cite{Braun1995OL}, since the critical power is inversely proportional to the gas pressure \cite{Couairon2006OC, Heyl2016Optica}. Therefore, in the case of HHG experiments carried out using low-pressure gas cells, beam trapping and guiding require higher intensity.
Still, the lasers used nowadays for HHG experiments can easily fulfill these higher-intensity requirements \cite{Kuhn2017JPB, Rivas2017SR}, and the intensity distribution in the interaction region can be determined by self-guiding.

At the same time, it is less often highlighted that for short pulses the rapid ionization can play a leading role in the guiding of a laser beam by the formation of a self-sustained plasma filament
\cite{Anderson1995PRE, Kim1996AIPCP, Sergeev1998Appl, Sergeev1999LPB, Rankin1991OL, Babin1996RQE, Decker1996PP,Miyazaki1995PRA}. Plasma-core induced beam self-guiding has been studied up to now in the context of beam smoothing \cite{Malka2003PRL}, frequency shifts \cite{Siders1996JOSAB} or other pulse propagation-related phenomena \cite{Bolton1996JOSAB, Decker1996PP}, but only a few times mentioned in relation to HHG in gases \cite{Miyazaki1995PRA, Tamaki1999PRL, Bellini2001PRA, Tosa2003PRA, Takahashi2003PRA, Vozzi2011NJP}.
In this work we present a numerical study on plasma-core induced self-guiding in connection with HHG in gases. 
Along with examining the properties and formation of this type of guiding, we also analyze the phase matching of the produced high-harmonic radiation in order to find the relation between the spatio-temporal structure of the laser field and the coherence length of the generated XUV.

This work is structured as follows. First, in Section \ref{sec:model}, we present the core equations of the three-dimensional non-adiabatic model that we use for the simulation of laser pulse propagation and HHG. We also introduce the expressions employed to analyze the spatio-temporal properties of phase matching. Then Section \ref{sec:results} is dedicated to the main results and their discussion in two parts: Section \ref{subsec:guiding} contains the description of the ionization-driven self-guiding, while Section \ref{subsec:PM} describes how the formation of the guided beam affects phase matching of HHG. Finally, in Section \ref{sec:concl} we summarize the most important conclusions. 

\section{Theoretical model}\label{sec:model}

We carried out our analysis using a three-dimensional non-adiabatic model described in detail in Refs. \cite{Tosa2009PRA,Tosa2016QE, Priori2000PRA}. 
Here we only summarize briefly the main steps of the simulation and highlight those aspects which are most important related to the current study.

The first step in the simulation is to calculate the propagation of the laser pulse $E_{l}(\vec{r},t)$ with central frequency $\omega_{l}$ in the medium by solving the non-linear wave equation \cite{Esarey1997IEEEJQE}
\begin{equation}\label{eq:waveeq}
\nabla^{2} E_{l}(\vec{r},t) - \frac{1}{c^{2}} \frac{\partial^{2} E_{l}(\vec{r},t)}{\partial t ^{2}} = \frac{\omega_{l}^{2}}{c^{2}} \left( 1 - \eta_{\mathrm{eff}}^{2}(\vec{r},t) \right) E_{l}(\vec{r},t)\,,
\end{equation}
where $c$ is the speed of light in vacuum.
The effective refractive index $\eta_{\mathrm{eff}}(\vec{r},t)$ is calculated as \cite{Tosa2016QE}
\begin{equation}\label{eq:refind}
\eta_{\mathrm{eff}}(\vec{r},t) = \eta_{0} + \bar{\eta}_{2} \langle E_{l}^{2}(\vec{r},t) \rangle - \frac{\omega_{p}^{2}(\vec{r},t)}{2 \omega_{l}}\,,
\end{equation}
where the angular brackets mean time averaging for an optical cycle and $\omega_{p}^{2} = n_{e} e^{2}/(m \varepsilon_{0}) $ is the square of the plasma frequency (expressed using the number density of electrons $n_{e}$, the elementary charge $e$, the mass of the electron $m$ and the vacuum permittivity $\varepsilon_{0}$). Dispersion and absorption ($\eta_{0}$), Kerr-effect ($\bar{\eta}_{2}$), plasma dispersion along with absorption losses due to ionization \cite{Geissler1999PRL} are taken into account. Our model assumes cylindrical symmetry ($\vec{r} \rightarrow (r,z)$) and uses the paraxial approximation \cite{Tosa2016QE}. By applying a moving frame with the speed of light $c$ and eliminating the temporal derivative using Fourier transform, the equation that is solved explicitly is
\begin{equation}\label{eq:wavefreq}
\begin{split}
\left( \frac{\partial^{2}}{\partial r^{2}} + \frac{1}{r} \frac{\partial}{\partial r} \right)\ &E_{l}(r,z,\omega) - \frac{2 i \omega}{c} \frac{\partial E_{l}(r,z,\omega)}{\partial z } = \\ &= \frac{\omega^{2}}{c^{2}} \mathcal{F} \left[ \left( 1 - \eta_{\mathrm{eff}}^{2} \right) E_{l}(\vec{r},t) \right]\,.
\end{split}
\end{equation}
The solution is obtained using the Crank-Nicolson method in an iterative algorithm \cite{Tosa2016QE}. The $\mathcal{F}$ symbol on the right-hand side of Eq. (\ref{eq:wavefreq}) means Fourier transform.
In all studied cases in this work a top-hat longitudinal pressure profile is assumed for the medium (a Gaussian pressure gradient can also be included), while the radial pressure distribution is constant. The laser field distribution in the input plane of the medium is calculated using the ABCD-Hankel transformation \cite{Ibnchaikh2001PCN}, based on the given pulsed beam and focusing geometry parameters.

The second step is the calculation of the single-atom response (dipole moment $x_{\mathrm{nl}}(t)$) based on the laser-pulse temporal shapes available on the full $(r,z)$ grid. This is carried out using the Lewenstein integral in the saddle-point approximation, expressed with atomic units \cite{Sansone2004PRA, Lewenstein1994PRA}
\begin{equation}
\begin{split}
&x_{\mathrm{nl}}(t)=2 \mathrm{Re} \left\{i\int_0^t \mathrm{d}t' \left[ \frac{\pi}{\epsilon + i(t-t')/2} \right]^{3/2} E_{l}(t') \right.\\ 
&\times D^{*}\! \left(p_{\mathrm{st}} - A_{l}(t)\right) \exp\!\left[-i S(p_{\mathrm{st}},t,t')\right] D\!\left(p_{\mathrm{st}} - A_{l}(t')\right) + \mathrm{c.c} \biggr\}\,,
\end{split}
\end{equation}
where $p_{\mathrm{st}} = \left(1/(t-t')\right) \int_{t'}^{t}A_{l}(t'')\,\mathrm{d}t''$ is the stationary canonical momentum, and $S(p,t,t')=\int_{t'}^{t} \left[ (p-A_{l}(t''))^{2}/2 + I_{p}\right] \mathrm{d}t''$ is the quasi-classical action with the vector potential of the laser field  $A_{l}(t)$, the ionization potential $I_{p}$ of the atom and $\epsilon$ is a regularization constant. The dipole matrix element $D$ is taken with a form corresponding to a hydrogen-like potential \cite{Lewenstein1994PRA}. For the macroscopic non-linear response $P_{\mathrm{nl}}(t)$ the depletion of the ground state is taken into account \cite{Lewenstein1994PRA, Le2009PRA}, giving
\begin{equation}
P_{\mathrm{nl}}(t) = n_{a} x_{\mathrm{nl}}(t) \exp\!\left[-\int_{-\infty}^{t} w(t')\,\mathrm{d}t' \right]\,,
\end{equation}
where $w(t)$ is the ionization rate calculated using the ADK model \cite{Ammosov1986JETP}, and $n_{a}$ is the number density of atoms in the specific grid point $(r,z)$ \cite{Gaarde2008JPB, Tosa2016QE}.

The third and last step is to calculate the propagation of the generated radiation $E_{h}(\vec{r},t)$ by the wave equation
\begin{equation}\label{eq:Hwaveeq}
\nabla^{2} E_{h}(\vec{r},t) - \frac{1}{c^{2}} \frac{\partial^{2} E_{h}(\vec{r},t)}{\partial t ^{2}} = \mu_{0} \frac{\mathrm{d}^{2} P_{\mathrm{nl}}(t)}{\mathrm{d}t^{2}}\,,
\end{equation}
$\mu_{0}$ being the vacuum permeability. Eq. (\ref{eq:Hwaveeq}) is solved similarly to Eq. (\ref{eq:waveeq}), but since the source term is known, there is no need for an iterative scheme. The dispersion and absorption of the harmonic field is also taken into account.

To analyze the phase matching properties we used the intuitive model of Balcou \emph{et al.} \cite{Balcou1997PRA}, which was extended to a time-dependent theory in Ref. \cite{Takahashi2003PRA}.
In this time-dependent concept the polarization wave vector of the harmonic order $q$ can be calculated as
\begin{equation}\label{eq:kq}
\vec{k}_{q} = q \nabla \left[ \phi_{l}(r,z) \right] + \nabla \left[ \alpha I_{l}(r,z,t) \right]\,,
\end{equation}
where $\phi_{l}(r,z)$ is the spectrally-averaged phase and $I_{l}(r,z,t)$ is the intensity (calculated from the envelope) of the laser field. Both previous quantities are obtained from the solution of Eq. (\ref{eq:waveeq}). The phase parameter
$\alpha$ is the well-known proportionality constant between the atomic phase and the intensity of the laser field \cite{Lewenstein1995PRA}.
With this approach the polarization wave-vector intrinsically contains non-geometrical effects, like dispersion, absorption and ionization. This means there is no need to take them into account separately like it is done in many cases \cite{Takahashi2003PRA, Tosa2003PRA, Vozzi2011NJP}, and it is unnecessary to make further assumptions on these different terms.
Finally, the coherence length is defined as
\begin{equation}\label{eq:Lcoh}
L_{\mathrm{coh}} = \frac{\pi}{\Delta k}, \quad \text{where} \quad \Delta k = \frac{q \omega_{l}}{c} - \left| \vec{k}_{q}  \right|\,.
\end{equation}
The above formulas allow one to analyze spatio-temporal phase matching characteristics based on the propagated laser field. Our three-dimensional simulation code and the phase matching expressions have been tested against experimental results successfully several times \cite{Tosa2003PRA, Takahashi2003PRA, Vozzi2011NJP, Rivas2018Optica}.

\section{Simulation results and discussion}\label{sec:results}

\subsection{Plasma-core induced self-guiding}
\label{subsec:guiding}

Plasma-core induced self-guiding has been studied in detail by Sergeev and co-workers in previous publications \cite{Anderson1995PRE, Kim1996AIPCP, Sergeev1999LPB, Babin1996RQE}. Here we first summarize the most important aspects previously published, and extend them with findings that are most important for HHG.

The key presumption of the model used by Sergeev \emph{et al.} is that ionization is considered as a threshold-type process with respect to the strength of the laser field \cite{Anderson1995PRE, Kim1996AIPCP}.
This is a fair assumption for tunneling ionization, since it depends highly nonlinearly on intensity. The ionization rate is thus given by
\begin{equation}
\gamma(E_{l}) =
\begin{cases}\label{eq:ionrate}
\infty & \quad \text{if } E_{l} \ge E_{\mathrm{th}} \\
0 & \quad \text{if } E_{l}<E_{\mathrm{th}}
\end{cases}\,,
\end{equation}
where $E_{\mathrm{th}}$ is a threshold electric field strength for ionization. This assumption is used in Ref. \cite{Anderson1995PRE} to obtain an analytical solution in a two-dimensional model.
The solution reads as
\begin{equation}\label{eq:guidedE}
E_{l}(x) =
\begin{cases}
E_{\mathrm{th}} & \quad \text{if } \lvert x \rvert \leq d \\
E_{\mathrm{th}}   \mathrm{sech} \left[ \Gamma (x-d) \right] & \quad \text{if } \lvert x \rvert > d
\end{cases}\,,
\end{equation}
where $\Gamma = E_{\mathrm{th}} \omega_{0} \eta_{0} \sqrt{ \eta_{2} \varepsilon_{0}}/\sqrt{2c}$ (for brevity of the expression $\eta_{2}$ -- defined by $\eta = \eta_{0} + \eta_{2} I$ -- is different from $\bar{\eta}_{2}$ appearing in Eq. (\ref{eq:refind}), see the relation of the two in Section 4.1 of Ref. \cite{Boyd}). In the two-dimensional case the plasma channel guiding the beam has a top-hat profile with $2d$ width \cite{Anderson1995PRE}. 

In three dimensions, assuming a cylindrical symmetry, only numerical solutions can be obtained even with the simplified ionization model of Eq. (\ref{eq:ionrate}). One such solution describes a radial laser field distribution $E_{l}(r)$ that has a constant amplitude equivalent to the threshold field strength $E_{\mathrm{th}}$ up to a radial distance $d$. For $r>d$, the electric field $E_{l}(r)$ changes according 
to the beam profile that was obtained for self-trapped optical beams more than half a century ago (the Townes profile \cite{Chiao1964PRL}). The radial distribution of electron density is similar to the Townes profile, so resembles the profile of the electric field \cite{Takahashi2003PRA}.

\begin{figure}[htbp]
	\begin{center}
		\includegraphics[width=0.7\linewidth]{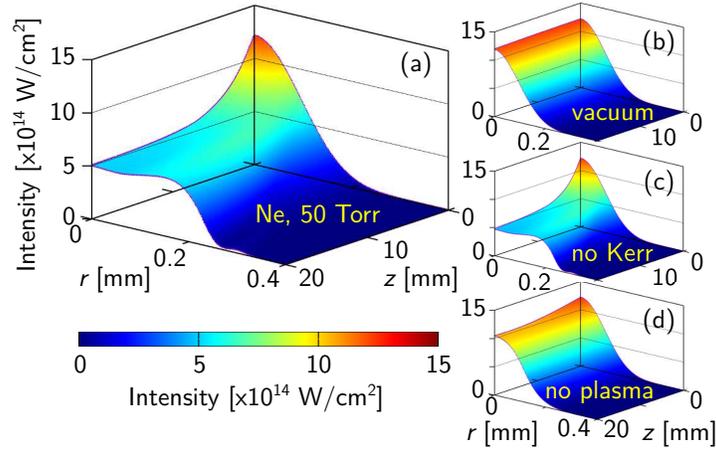}
		\caption{(a) The peak intensity distribution of a pulsed laser beam focused into $50\,\mathrm{Torr}$ Ne gas medium. The laser pulse is $15\,\mathrm{fs}$ long with Gaussian temporal shape, $800\,\mathrm{nm}$ central wavelength and $10\,\mathrm{mJ}$ energy. In the simulation a Gaussian beam of $15\,\mathrm{mm}$ $\mathrm{1/e^2}$ radius is assumed to be focused by a $10\,\mathrm{m}$ focal length focusing element and the gas cell begins in the geometrical focus. Transverse profile of the same pulsed beam focused into (b) vacuum, (c) into Ne without considering the Kerr contribution and (d) into Ne neglecting the effect of plasma on the refractive index. Absorption is included in all cases except (b).}
		\label{fig:guiding_comps}
	\end{center}
\end{figure}

As a representative case, the formation of a beam profile similar to that of Eq. (\ref{eq:guidedE}), obtained with our three-dimensional non-adiabatic HHG simulation code can be seen in Fig. \ref{fig:guiding_comps}(a). The figure clearly shows that the intensity of the laser beam that is focused into a gaseous medium is self-regulated to a value which can be found to be the threshold intensity $I_{\mathrm{th}}$ corresponding to the ionization threshold electric field $E_{\mathrm{th}}$ in Eq. (\ref{eq:ionrate}). This intensity was also found previously as the intensity of the trapped pulse for the Townes mode \cite{Braun1995OL}.

To find the reason for the guiding, we carried out simulations where we turned off different contributions to the refractive index. Fig. \ref{fig:guiding_comps} shows the evolution of the beam profile in vacuum (b), when no Kerr effect is included (c),
and when no plasma formation is considered (d) in Eq. (\ref{eq:refind}). Absorption is also included in (a), (c) and (d). In vacuum (see Fig. \ref{fig:guiding_comps}(b)), there is almost no change in the beam profile during the $20\,\mathrm{mm}$ propagation, since the Rayleigh length of the focused beam is more than three times longer ($\sim 76\,\mathrm{mm}$) than the propagation length. Without the Kerr term (see Fig. \ref{fig:guiding_comps}(c)), the change of beam profile is very similar to the situation when all terms are included in the expression of the refractive index (cf. Fig. \ref{fig:guiding_comps}(a)). When the effect of free electrons is excluded (see Fig. \ref{fig:guiding_comps}(d)), the changes resemble  those in vacuum (cf. Fig. \ref{fig:guiding_comps}(b)), except a slight continuous decrease of peak intensity during propagation due to absorption. 
So, the results depicted in Fig. \ref{fig:guiding_comps} clearly show that ionization has a dominant role in the formation of the guided structure. This can be explained as follows. The leading part of the pulse generates a plasma-core with a sharp refractive index gradient which serves as a waveguide for the remaining part of the pulse. The observation that ionization has a dominant role in this type of guiding is in agreement with the previous finding that qualitatively the same steady-state results can be obtained when no focusing nonlinearity is considered in the model ($\eta_{2} = \bar{\eta}_{2}=0$) \cite{Kim1996AIPCP,Gildenburg1975JETP}. This type of laser guiding \cite{Takahashi2003PRA}, also called ``ionization-induced leaking-mode channeling'' in previous work \cite{Sergeev1999LPB, Sergeev1998Appl},
has several interesting features and consequences on HHG. In the following few paragraphs we present the results of several numerical simulations, which were carried out with different input conditions. The large variety of conditions aims to investigate the generality of our conclusions. 

\begin{figure}[htbp]
	\begin{center}
		\includegraphics[width=0.7\linewidth]{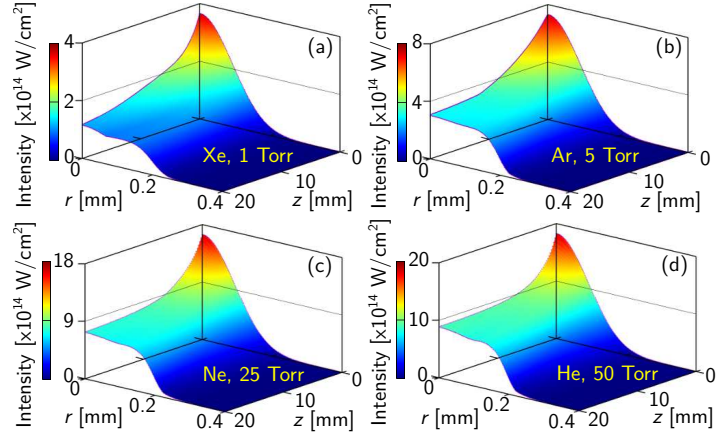}
		\caption{The evolution of the spatial beam profile of a $5\,\mathrm{fs}$, $800\,\mathrm{nm}$ pulsed beam when focused into gaseous media of different type and pressure: (a) $1\,\mathrm{Torr}$ Xe, (b) $5\,\mathrm{Torr}$ Ar, (c) $25\,\mathrm{Torr}$ Ne and (d) $50\,\mathrm{Torr}$ He. The pulse energies are (a) $1\,\mathrm{mJ}$, (b) $2\,\mathrm{mJ}$, (c) $4.5\,\mathrm{mJ}$ and (d) $5\,\mathrm{mJ}$. Other focusing and pulsed beam parameters are the same as in Fig. \ref{fig:guiding_comps}.
		}
		\label{fig:guiding_gas}
	\end{center}
\end{figure}

As a first feature, the peak intensity to which the beam is regulated during propagation depends almost only on the used gas type. This regulated threshold intensity is, e.g.,
$I_{\mathrm{th}}^{\mathrm{Xe}} \approx (0.8-1)\cdot10^{14}\,\mathrm{W/cm^{2}}$ for Xe, $I_{\mathrm{th}}^{\mathrm{Ar}} \approx (2-3)\cdot10^{14}\,\mathrm{W/cm^{2}}$ for Ar, $I_{\mathrm{th}}^{\mathrm{Ne}} \approx (6-8)\cdot10^{14}\,\mathrm{W/cm^{2}}$ for Ne, and $I_{\mathrm{th}}^{\mathrm{He}} \approx (9-11)\cdot10^{14}\,\mathrm{W/cm^{2}}$ for He.
These threshold intensities are in accordance, e.g., with the experimentally obtained values of Augst \emph{et al.} \cite{Augst1989PRL} termed there as ``ion appearance intensities'', later introduced as barrier-suppression intensities \cite{Scrinzi1999PRL}.  Fig. \ref{fig:guiding_gas} depicts the evolution of the laser beam profile in these gases and the regulation of peak intensity to the values mentioned above. In the subfigures of Fig. \ref{fig:guiding_gas}, the pressures are different for each gas type chosen to optimize the formation of a plasma channel since a lower ionization probability means that a higher number of atoms is necessary for the formation of a plasma channel that has high enough electron density to guide the beam. The pulse energy of the laser beam is also different in each case, because when the peak intensity of the beam entering the medium is much higher than the threshold intensity $I_{\mathrm{th}}$ of the specific gas, plasma defocusing is too strong to allow for the formation of the guided beam profile. 

\begin{figure}[htbp]
	\begin{center}
		\includegraphics[width=0.7\linewidth]{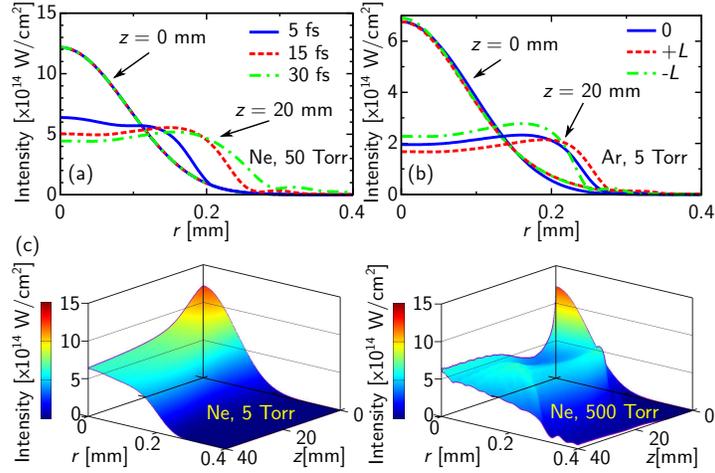}
		\caption{(a) Initial and propagated beam profiles in Ne for different pulse durations. The simulation parameters are otherwise the same as in Fig. \ref{fig:guiding_comps}, except energies, which were tuned to yield the same focused beam profile at the entrance of the medium ($z=0\,\mathrm{mm}$). (b) Initial and propagated beam profiles in Ar with different wave-front curvatures. The cell was assumed to begin in focus (0), and at one Rayleigh length ($L = \omega_{l} w_{0}^{2}/(2c)$) distances from focus ($\pm L$). The focal length and pulse energy were tuned to yield approximately the same input intensity profile ($z=0\,\mathrm{mm}$) to minimize the effect of differences in intensity distribution. (c) Evolution of beam profiles in Ne with different pressures ($5\,\mathrm{Torr}$ and $500\,\mathrm{Torr}$). Apart from pressure the simulation parameters are the same as in Fig. \ref{fig:guiding_comps}, so Fig. \ref{fig:guiding_comps}(a) shows the result for a third, intermediate pressure value ($50\,\mathrm{Torr}$).}
		\label{fig:guiding_deps}
	\end{center}
\end{figure}

Another consequence of the highly nonlinear intensity dependence of ionization is that the variation of the regulated intensity with pulse duration or central wavelength is limited (for variation with pulse duration see Fig. \ref{fig:guiding_deps}(a)). This is also suggested by the simplified ionization model of the analytical theory, since the threshold intensity $I_{\mathrm{th}}$ only slightly varies with pulse duration or central wavelength.
Also, the shape of the beam formed during the regulation of the intensity is very similar regardless of the wave-front curvature or the focus position with respect to the cell. This can be seen in Fig. \ref{fig:guiding_deps}(b), where the beam and focusing parameters were tuned to give very similar intensity profiles but different wave-fronts at the entrance of the Ar medium, in order to separate the effects of intensity profile from those of the wave-front shape. In the three depicted cases of Fig. \ref{fig:guiding_deps}(b), the cell begins in the focus, and at one Rayleigh length ($L = \omega_{l} w_{0}^{2}/(2c)$, $w_{0}$ being the beam waist radius) from it. Throughout the paper, positive cell positions mean that the medium is placed after the focus, while negative values mean a cell before the focus.  

While the beam profile is similar regardless of the wave-front curvature or medium position with respect to laser focus, the guided distance depends strongly on input parameters (see later in Section \ref{subsec:PM}).
The simplified analytical model summarized above totally neglects losses allowing for beam guiding on an infinite propagation length  \cite{Kim1996AIPCP, Anderson1995PRE}. In real situations, absorption and energy flux across the plasma boundary limits the guiding length.

The pressure of the gas, e.g., influences the length of the plasma channel (see Fig. \ref{fig:guiding_deps}(c)). The reason is that the medium needs to produce enough electrons for the plasma-core to be formed \cite{Anderson1995PRE}.
The pressure also affects how fast the regulation is reached, as it can be seen in Fig. \ref{fig:guiding_deps}(c). In the $5\,\mathrm{Torr}$ case (left plot of Fig. \ref{fig:guiding_deps}(c)), the low pressure is just enough to support the plasma channel and the formation of a steep radial gradient in refractive index profile. Upon longer propagation, the beam simply gets diffracted not reaching steady-state, showing a continuous slow drop of intensity along the propagation direction. This slow decrease of intensity can be counteracted by focusing. The minimum pressure needed for clear guiding depends on the gas type due to the different ionization rates. In the $500\,\mathrm{Torr}$ case (see right plot of Fig. \ref{fig:guiding_deps}(c)), the peak intensity is reduced after a short distance below the threshold intensity, and a field structure similar to Eq. (\ref{eq:guidedE}) is not formed. The beam shape is heavily distorted. Still, at both pressures of Fig. \ref{fig:guiding_deps}(c) (and at other intermediate pressure values) the intensity is regulated to a value close to that predicted by the plasma-core induced guiding theory, i.e., close to the barrier-suppression intensity.
The top-hat-like beam profile and the regulated intensity value is reached also if we assume a longitudinal pressure gradient at the two ends of the cell (even when the gradient has the same length as the constant-pressure region).
It is also a general observation in the simulations that the pulsed laser beam is guided along a longer propagation length if the medium is placed
before the focal point, since this way geometrical focusing helps to maintain the beam shape by acting against diffraction effects.

In conclusion of this study, in (long) gas cells with the usual focused intensity and pressure range for gas HHG the amplitude of the field is regulated to a value that is primarily defined by the used gas type. The beam shape reached during this intensity regulation is determined by the intensity distribution at the entrance to the cell (the focusing geometry), while not much affected by the spatial phase properties. 
A guided top-hat-like electric field structure can be formed in which the peak intensity corresponds to the ionization threshold intensity of the medium.

\subsection{Phase matching of HHG in plasma-core induced guiding}
\label{subsec:PM}

\begin{figure}[htbp!]
	\begin{center}
		\includegraphics[width=0.7\linewidth]{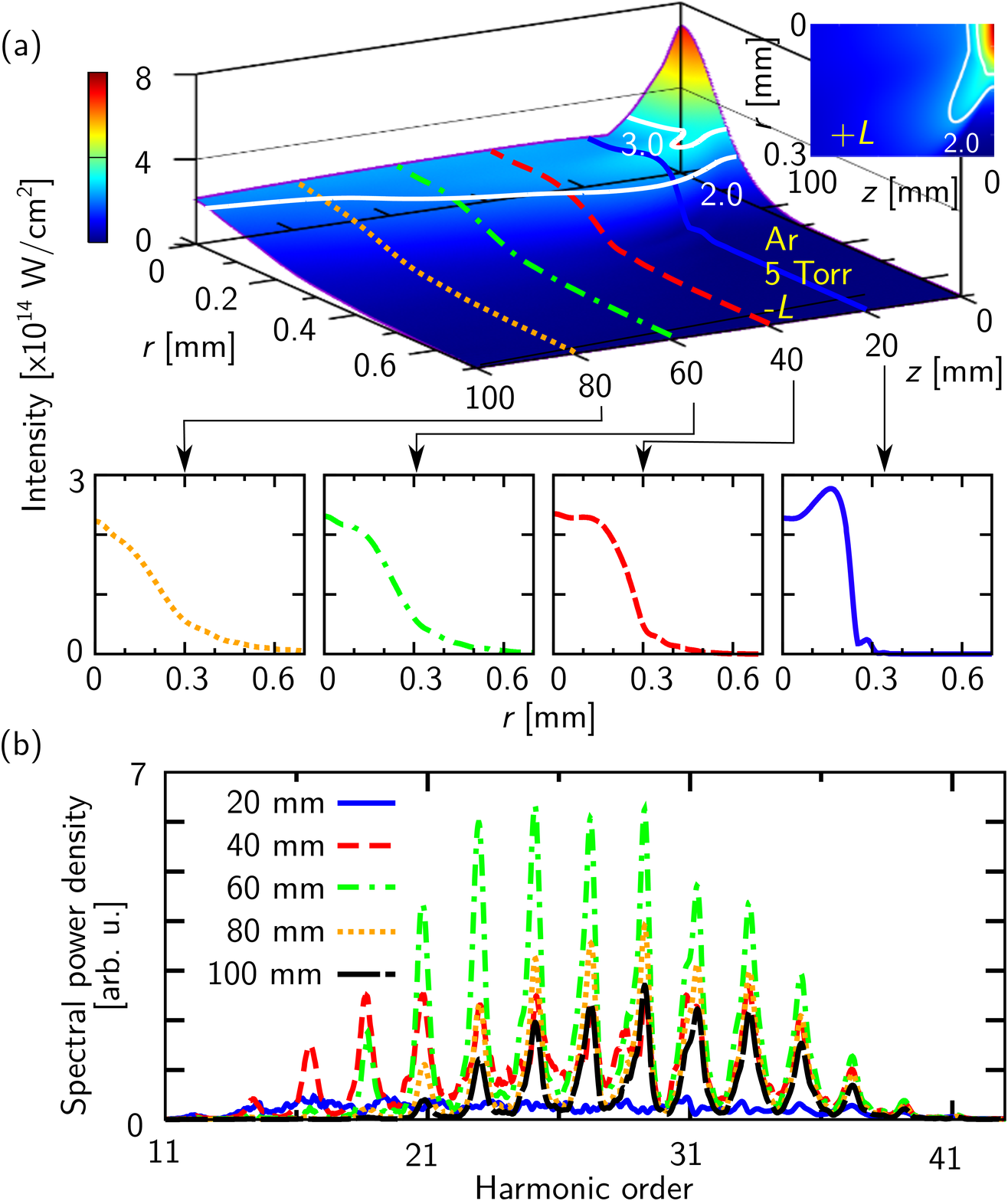}
		\caption{(a) The laser beam peak intensity profile evolution in $5\,\mathrm{Torr}$ Ar. A $10.7\,\mathrm{mJ}$, $27\,\mathrm{fs}$, $800\,\mathrm{nm}$ laser pulse with a 
		beam radius of $15\,\mathrm{mm}$ ($1/e^{2}$) served as the input. The beam is focused with a $8.2\,\mathrm{m}$ focal length focusing element, and the gas cell is positioned $7.6\,\mathrm{cm}$
		before the focus (one Rayleigh length of the central wavelength, $-L$). The middle plots show the beam profiles at $20, 40, 60\text{ and }80\,\mathrm{mm}$ propagation distances.
		The inset in the top right corner depicts the peak intensity evolution when the cell is placed one Rayleigh length behind focus ($+L$).
		The white lines highlight the spatial contour of $2$ and $3\cdot10^{14}\,\mathrm{W/cm^{2}}$ peak intensity. 
		(b) The evolution of the spatially integrated harmonic signal along the cell.}
		\label{fig:diffpos}
	\end{center}
\end{figure}

We now study phase matching in relation with plasma-core induced self-guiding. 
The guided length is influenced by both the intensity profile and the phase properties of the input beam or by the relative position of the medium with respect to the laser focus (see Fig \ref{fig:diffpos}(a)).
For the analysis of phase matching, we choose a case where beam guiding is maintained for a longer propagation length,
which can be achieved by positioning the gas medium before the geometrical focus.  
Such a situation is depicted in Fig. \ref{fig:diffpos}(a), where
the simulated propagation length is long enough for the guiding to break off, making it possible to identify which of the observed features are related to guiding.

When the cell is placed after the geometrical focus, the flat-top beam profile and ionization-driven guiding is formed, but with the additional losses
due to diffraction the guided beam length is shorter.
The influence of focusing on maintaining longer guiding is shown by the fact that when the cell is placed behind the focus (see the inset of Fig. \ref{fig:diffpos}(a))
--- so in a diverging beam --- the extent of guiding is much shorter. The guided length in the medium placed after focus is a few millimeters compared to the $\sim 60\,\mathrm{mm}$ which is obtained when the cell begins before focus (see the main plot of Fig. \ref{fig:diffpos}(a)).
Also, we note that the end of the guided region is in the vicinity of the geometrical focal point when the cell is placed before focus, which is $7.6\,\mathrm{cm}$ away from the beginning of the cell,
underlying again the role of focusing.

In Fig. \ref{fig:diffpos}(b) one can see the spectral power density of the harmonic signal in the planes
highlighted by the cutouts of Fig. \ref{fig:diffpos}(a) and also at the end of the $10\,\mathrm{cm}$ cell. It is clear from the power spectra that the highest HHG signal
is reached approximately at the end of the guided region (the white contours in Fig. \ref{fig:diffpos}(a) aims to give a hint on the extent of guiding).
The above suggest that the spatially constant intensity provided by guiding offers favorable phase matching conditions for harmonics below the cut-off defined by
the regulated intensity level (the barrier-suppression intensity of $2.5\cdot 10^{14}\,\mathrm{W/cm^{2}}$ gives a cut-off of $\sim37$th order in the case of Ar) \cite{Tosa2003PRA}.

\begin{figure}[htbp!]
	\begin{center}
		\includegraphics[width=0.7\linewidth]{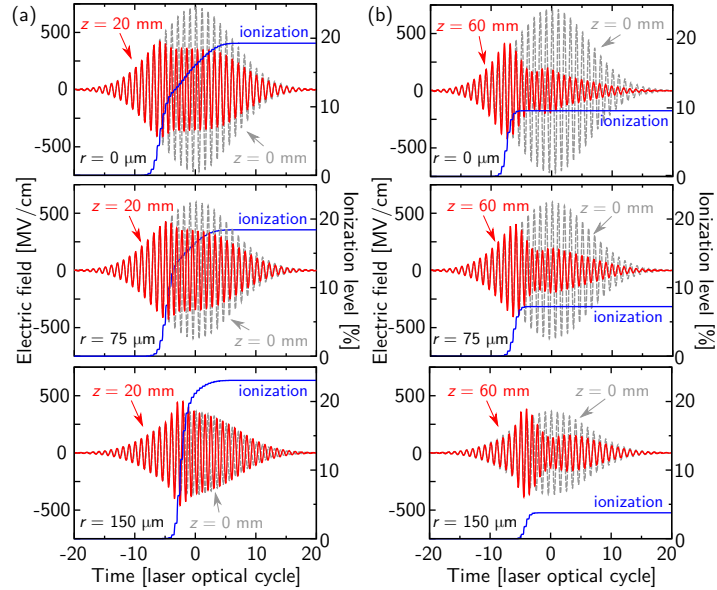}
		\caption{Temporal profiles of the laser electric field at different radial coordinates after (a) $20\,\mathrm{mm}$ and (b) $60\,\mathrm{mm}$ propagation in $5\,\mathrm{Torr}$ Ar medium (red curves).
		Further laser and focusing parameters can bee found in the caption of Fig. \ref{fig:diffpos}. The ionization ratios (blue curves) are also depicted. 
		In each plot the gray dashed curve shows the temporal profile at the beginning of the medium ($z=0\,\mathrm{mm}$) at the same radial coordinate for comparison.}
		\label{fig:fieldtemporal}
	\end{center}
\end{figure}

There is a further peculiarity of ionization-induced self guiding for longer pulses, which is highlighted in Fig. \ref{fig:fieldtemporal}.
The intensity stabilizes in the guided region not jut spatially but also temporally in the middle of the pulse (at $0T$ of the moving temporal frame, 
$T$ being the optical cycle).
Fig. \ref{fig:fieldtemporal}(a) depicts the temporal profile of the laser electric field strength after $20\,\mathrm{mm}$
propagation in the Ar medium at three different radial coordinates ($0, 75 \text{ and } 150\,\text{\textmu m}$).
The $150\,\text{\textmu m}$ radial distance from axis means approximately the radial coordinate where there is a huge drop in intensity of 
the flat-top-like guided beam profile (see Fig. \ref{fig:diffpos}(a)).
Up to this radial distance from propagation axis, the envelope of the electric field is temporally steady in the window between $-2.5$ and $+2.5 T$, while
at $150\,\text{\textmu m}$ (and further from axis) this property is lost (see bottom plot of Fig. \ref{fig:fieldtemporal}(a)).

With further propagation (see Fig. \ref{fig:fieldtemporal}(b)) the spatial region with temporally stabilized field envelope is radially increased, but at the same time the pulse shows signatures of splitting due to the longer propagation in the highly ionized medium \cite{Tosa2016QE} (see ionization levels 
in Fig. \ref{fig:fieldtemporal} with blue continuous curves). This effect is stronger on axis,
resulting in more stable temporal envelope at off-axis coordinates in the case of longer propagation. 
As can be seen both in Fig. \ref{fig:diffpos}(a) and in Fig. \ref{fig:fieldtemporal},
the peak intensity is approximately the same
after $20\,\mathrm{mm}$ and $60\,\mathrm{mm}$ propagation.
At the same time, the amplitude of the envelope in the temporally constant domain is substantially decreased
compared to the situation with shorter propagation distance (cf. Fig. \ref{fig:fieldtemporal}(a) with (b)). This limits the achievable photon energy
of harmonics generated in these optical cycles.

\begin{figure}[htbp!]
	\begin{center}
		\includegraphics[width=0.7\linewidth]{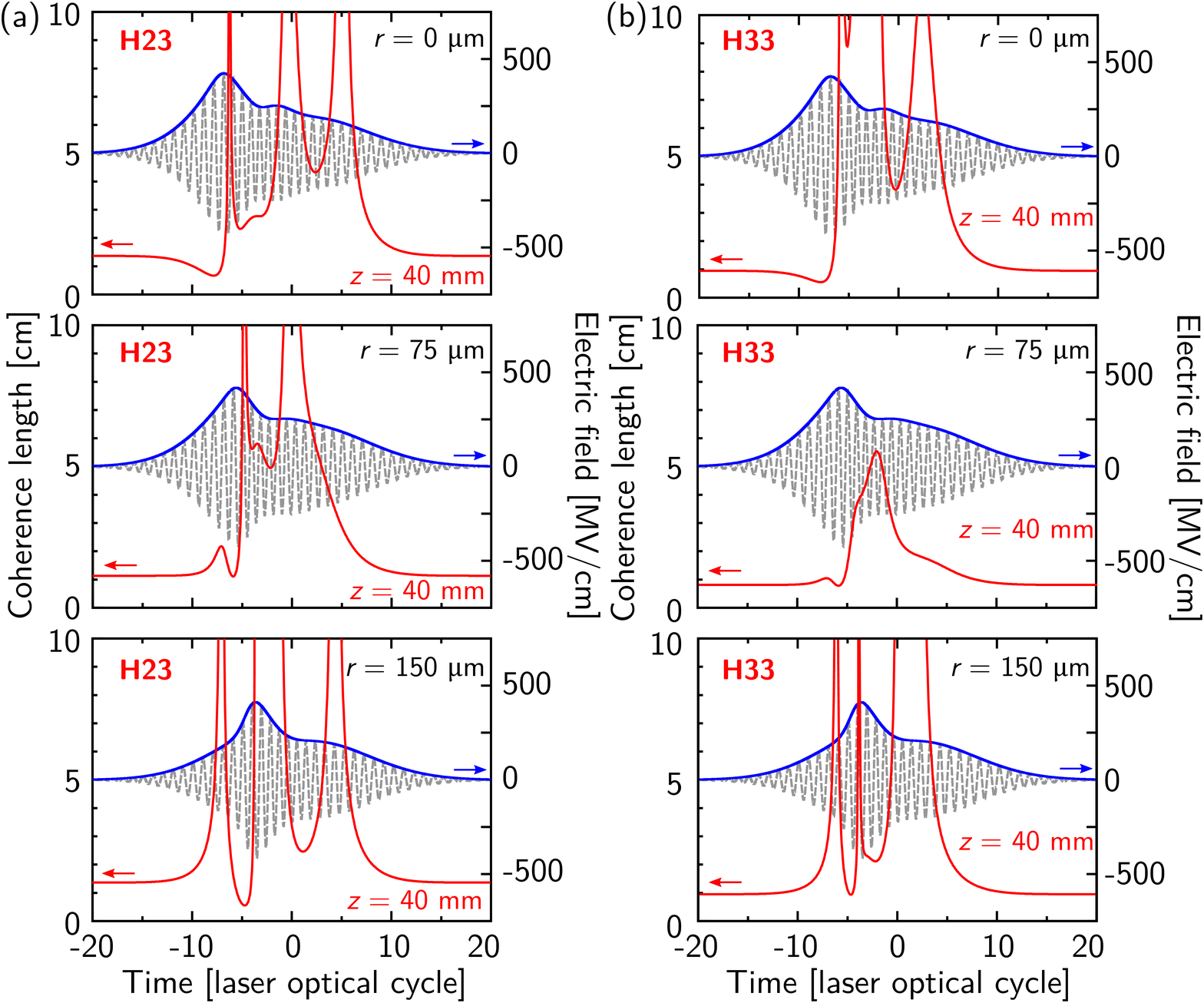}
		\caption{The temporal variation of the coherence length for the (a) 23rd and (b) the 33rd harmonic orders after $40\,\mathrm{mm}$ propagation in $5\,\mathrm{Torr}$ Ar gas
		at different radial coordinates.
		The laser and focusing parameters are the same as in the caption of Fig. \ref{fig:diffpos}. The coherence length was obtained using Eq. (\ref{eq:Lcoh}).
		The electric field strengths and envelopes are also depicted with dashed gray and blue continuous curves, respectively.}
		\label{fig:cohlength}
	\end{center}
\end{figure}

To analyze the phase matching conditions, we plotted the temporal dependence of the coherence length for different harmonic orders after $40\,\mathrm{mm}$ 
propagation (see Fig. \ref{fig:cohlength}). $40\,\mathrm{mm}$ propagation distance was chosen because it is in between the onset of guiding ($\sim 20\,\mathrm{mm}$) and the position where guiding slowly starts to break off ($\sim 60\,\mathrm{mm}$).
By comparing the time-dependent coherence length of the 23rd harmonic (Fig. \ref{fig:cohlength}(a)) with that of the 33rd harmonic (Fig. \ref{fig:cohlength}(b))
it can be seen that in both cases these harmonic orders are best phase matched in the time domain where the laser field's envelope has a plateau
(mostly between $-2.5$ and $+2.5$ optical cycles), or has a saddle-point-like temporal variation. At the same time the coherence length properties depend considerably on harmonic order.

In the studied cases the following tendencies were observed: 
(i) higher orders are phase matched at longer propagation lengths, 
(ii) upon further propagation the volume of highest coherence length
spreads radially, while the optimal region (meaning longest time duration with high coherence length) gets off axis. 
So the volume of optimal phase matching for each harmonic order can be imagined as the region that a diverging annular beam would illuminate while
propagating through the volume. 

The off-axis-shifted optimum can be seen in Fig. \ref{fig:cohlength}(a), where for the 23rd harmonic phase matching at $r = 75\,\text{\textmu m}$
is better than at $r = 0\,\text{\textmu m}$. This optimum is closer to the axis for the 33rd harmonic (cf. Fig. \ref{fig:cohlength}(b)),
and the radial extent of long coherence length generation is smaller compared to that of the 23rd harmonic. We thus conclude that in the presence of ionization-induced guiding, higher harmonic orders are generated with a long coherence length in a smaller volume compared to lower orders. It is to be noted that although long-duration phase matching
happens off-axis, long coherence length radiation is generated almost along the whole cross section of the guided beam. 
Such off-axis optimum for phase matching in the case of media placed before the focus 
agrees with earlier results in conditions of weak ionization \cite{Salieres1995PRL}.
However, with the higher ionization levels in the present study (see ionization levels in Fig. \ref{fig:fieldtemporal}), 
phase matching optimum can be found off-axis also for cells behind the focal point.
This is because similar guided beam profiles are formed irrespective of cell positions.
However, when the cell is behind the focus the length of guiding is much shorter.

\begin{figure}[htbp]
	\begin{center}
		\includegraphics[width=0.7\linewidth]{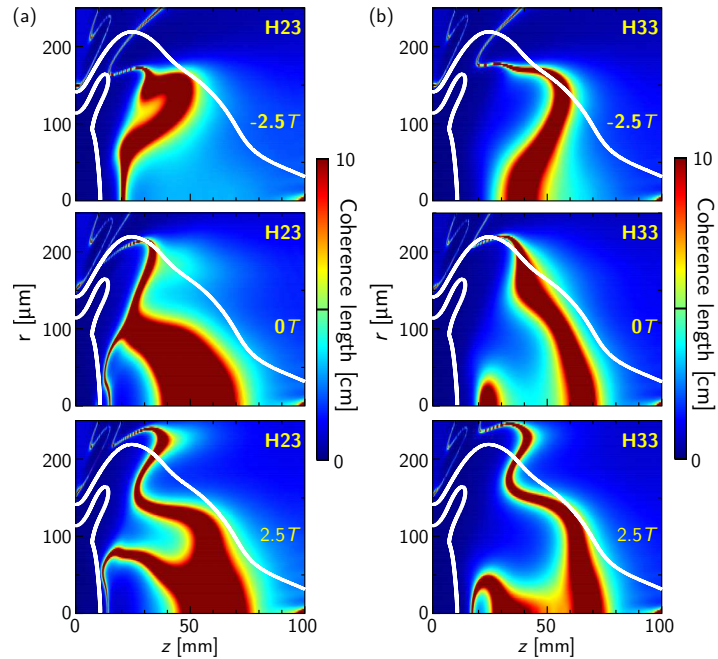}
		\caption{Spatial variation of the coherence length for the (a) 23rd and (b) 33rd harmonic at different time instants (at $-2.5, 0 \text{ and } +2.5$ optical cycles, covering the temporal window of long coherence length (see Fig. \ref{fig:cohlength}). The coherence lengths are calculated using Eq. (\ref{eq:Lcoh}). The generation parameters are the same as in Fig. \ref{fig:diffpos}.
		The white curves show the contour of $2$ and $3\cdot 10^{14}\,\mathrm{W/cm^{2}}$ peak intensity, aiming to show the guiding volume like in Fig. \ref{fig:diffpos}(a).}
		\label{fig:PMmap}
	\end{center}
\end{figure}

To highlight the correlation between the guided region and the well phase matched interaction volume for HHG, in Fig. \ref{fig:PMmap} we plot the coherence length as a function
of position in a $10\,\mathrm{cm}$ medium at different time instants.
In case of the 23rd harmonic (Fig. \ref{fig:PMmap}(a)), if one puts the plots obtained at different time instants ($-2.5T, 0T \text{ and } +2.5 T$) on top of each other, the long coherence length regions ($L_{\mathrm{coh}}>10\,\mathrm{cm}$) together overlap with the guided region (area between the white contour lines in Fig. \ref{fig:PMmap}). Carrying out the same ``projection'' with the phase matching maps of the 33rd harmonic (plots of Fig. \ref{fig:PMmap}(b)) leads to a similar conclusion: harmonics generated in the guided volume have longest coherence lengths.
Taking phase matching maps outside the constant-envelope temporal domain (before $-2.5T$ and after $2.5T$ in this case) gives distributions with short coherence lengths.
Further analysis with different generation conditions also showed similar correlation between phase matching and guiding.
For a cell placed after the focus, e.g., high-coherence length radiation is generated only up to the first $20\,\mathrm{mm}$ of propagation,
in correspondence with the intensity distribution and guided length (see the intensity profile for this situation in the inset of Fig. \ref{fig:diffpos}(a)). 

Overall, best phase matching is achieved when the intensity of the generating fundamental field is constant both spatially
and temporally. At the same time, it should be taken into account that even though phase matching is good in the whole guided region,
with longer propagation length the intensity is decreased during the temporal plateau of the laser field, so the cut-off is decreased, and higher orders are generated less efficiently.

\section{Conclusions}\label{sec:concl}

We analyzed ionization-induced self-guiding and its effect on HHG in gases. We showed that under usual generation conditions (gas pressures, focusing geometries, pulse durations \dots etc.) for HHG, self-guiding can be formed and maintained for long propagation lengths in gas cells placed before the laser focus. An important consequence of the formation of the guided beam structure is that the intensity is regulated to a value close to the barrier-suppression intensity. This regulated intensity level appears to depend practically only on the used gas type, while being almost independent from other conditions (gas pressure, wave-front curvature, gas cell position, pulse duration, central wavelength).

We also studied phase matching of high-order harmonics in conditions of ionization-induced self-guiding. The constant pulsed laser intensity both spatially and temporally in the guided region of the interaction volume provides favorable phase matching conditions. Overall, the general role that ionization-induced self-guiding can play in phase matched generation of high-order harmonics in gases,
being a primary source for high-flux attosecond sources, was highlighted. It was shown that placing the generation medium before the geometrical focus can extend
the guided length and as such the phase matched volume and the high-harmonic flux and is thus favorable for building high power high-harmonic sources.

\section*{Funding}
European Regional Development Fund (GINOP-2.3.6-15-2015-00001), Romanian National Authority for Scientific Research and Innovation (RO-CERN 03ELI (PROPW)), Swedish Research Council, European Research Council (Grant PALP 339253), Knut and Alice Wallenberg Foundation.

\section*{Acknowledgement}
We acknowledge KIF\"U for awarding us access to supercomputing resources based in Hungary at Debrecen and at Szeged. We are also grateful for the access to the Data Center at INCDTIM Cluj-Napoca.


\bibliography{manuscript_working_intensity_references}

\end{document}